\documentclass{iau}
\usepackage{graphicx}
\usepackage{url}

\title[] 
{The Milky Way's halo in 6D: {\it Gaia}'s Radial Velocity Spectrometer performance}

\author[George Seabroke et al.]   
{George Seabroke$^{1}$, Mark Cropper$^1$, David Katz$^2$, Paola Sartoretti$^2$, Pasquale Panuzzo$^2$, Olivier Marchal$^2$, Alain Gueguen$^2$, Kevin Benson$^1$, Chris Dolding$^1$, Howard Huckle$^1$, Mike Smith$^1$, Steve Baker$^1$}

\affiliation{
$^1$Mullard Space Science Laboratory, University College London, Dorking, Surrey, UK\\
$^2$Observatoire Paris-Site de Meudon, GEPI, Paris, France}

\pubyear{2015}
\volume{317}  
\pagerange{119--126}
\jname{The General Assembly of Galaxy Halos}
\editors{A. Bragaglia, M. Arnaboldi, M. Rejkuba, D. Romano, eds.}
\begin{document}

\maketitle

\begin{abstract}
{\it Gaia}'s Radial Velocity Spectrometer (RVS) has been operating in routine phase for over one year since initial commissioning. RVS continues to work well but the higher than expected levels of straylight reduce the limiting magnitude.  The end-of-mission radial-velocity (RV) performance requirement for G2V stars was 15 km~s$^{-1}$ at $V = 16.5$ mag.  Instead, 15 km~s$^{-1}$ precision is achieved at $15 < V < 16$ mag, consistent with simulations that predict a loss of 1.4 mag.  Simulations also suggest that changes to {\it Gaia}'s onboard software could recover $\sim$0.14 mag of this loss. Consequently {\it Gaia}'s onboard software was upgraded in April 2015. The status of this new commissioning period is presented, as well as the latest scientific performance of the on-ground processing of RVS spectra.  We illustrate the implications of the RVS limiting magnitude on {\it Gaia}'s view of the Milky Way's halo in 6D using the {\it Gaia} Universe Model Snapshot (GUMS).
\end{abstract}
\firstsection 
              \vspace*{-0.5 cm}
\section{{\it Gaia}'s Radial Velocity Spectrometer performance}
\label{sec1}

{\it Gaia}'s Radial Velocity Spectrometer \cite{cropper2011} data are processed on-ground by the Data Processing and Analysis Consortium (DPAC) Co-ordination Unit (CU) 6 Spectroscopic Processing pipeline \cite{katz2011}.  The pipeline formally runs at the CU6 Data Processing Centre.  We present offline tests of the CU6 pipeline running at the Mullard Space Science Laboratory.  As already presented in \cite{cropper2014},  Fig. \ref{fig1} (left) presents tentative evidence that the CU6 pipeline is able to achieve the end-of-mission RV precision predicted by simulations that include the straylight.  The majority of the stars in Fig. \ref{fig1} (left) are G dwarfs so the measured $\sigma_{RV} \sim 15$ km~s$^{-1}$ at $15 < V < 16$ mag is consistent with simulations that predict a loss of 1.4 mag.\footnote{\url{http://www.cosmos.esa.int/web/gaia/science-performance}}

 \vspace*{-0.5 cm}
\section{The Milky Way's halo in 6D}

\begin{figure}[b]
\begin{center}
\includegraphics[width=0.49\textwidth]{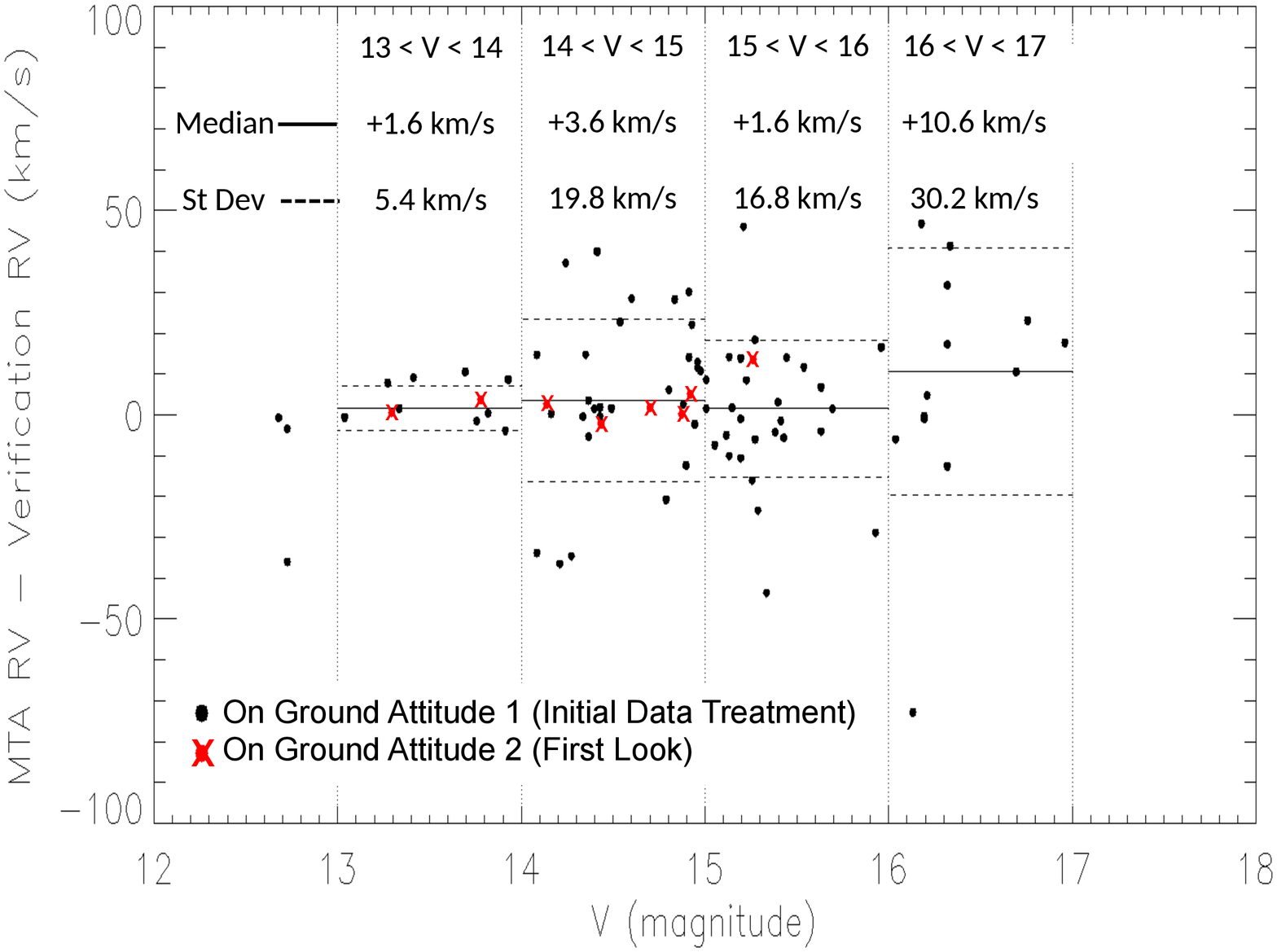} 
 \includegraphics[width=0.41\textwidth]{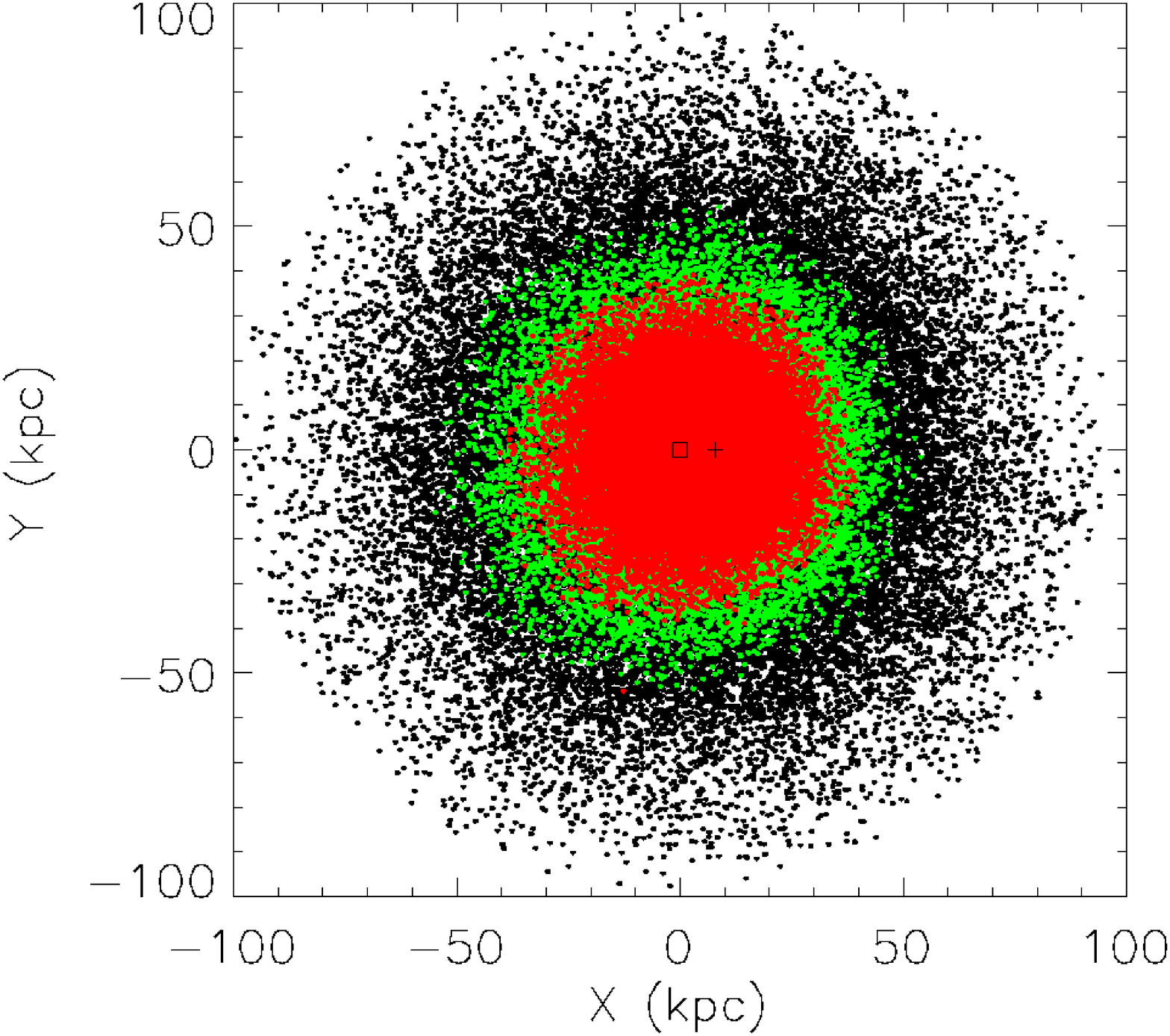} 
 \caption{{\it Left}: Measured end-of-mission (40 transits) RV performance as a function of magnitude.  {\it Right}: GUMS simulation of Milky Way halo stars with $G_{RVS} < 16.2$ mag looking face on: X and Y are in the Galactic plane with the Sun at the origin (square), where the cross is the Galactic centre. Each star is colour-coded according to its predicted end-of-mission $\sigma_{RV}$, assuming every star is a metal-poor K1 giant: $\sigma_{RV} < 15$ km~s$^{-1}$ (red/dark grey), $15 < \sigma_{RV} < 30$ km~s$^{-1}$ (green/light grey), $\sigma_{RV} > 30$ km~s$^{-1}$ (black).}
\label{fig1}
\end{center}
\end{figure}

The fixed limiting magnitude of $G_{RVS} = 16.2$ mag onboard RVS was changed to an adaptive one in September 2015.  Now the limiting magnitude ranges from $15.5 < G_{RVS} < 16.2$ mag as a function of time, corresponding to the straylight pattern to save telemetry on spectra that contain more noise than signal.  Selecting Milky Way halo stars from the GUMS simulation \cite{robin2012} with $G_{RVS} < 16.2$ mag suggests that RVS is collecting spectra for $\sim$1.8 million stars in the Milky Way's halo, the majority of which will be metal-poor giants.  Having approximately verified the RV performance model in Section \ref{sec1} with preliminary measurements, we apply this model to each GUMS star.  This predicts an {\it upper} limit (due to the adaptive limiting magnitude) on the number of halo stars in the final {\it Gaia} catalogue with $\sigma_{RV} < 15$ km~s$^{-1}$: $\sim$800,000, $\sim$42\% of the observed halo stars (red/dark grey dots in Fig. \ref{fig1} right).  Fig. \ref{fig1} (left) presents tentative evidence that $\sigma_{RV}$ can be measured up to 30 km~s$^{-1}$: $\sim$1.2 million stars, 67\% of the observed halo stars (green/light grey dots in Fig. \ref{fig1} right). Beyond this limit, it has not been tested that the data follow Poisson statistics and so whether the predictions are valid.

RVS spectra were obtained with a fixed ACross Scan (AC) window width of 10 pixels. The RVS S/N simulator suggests that adapting the AC width to the observing conditions could mitigate the impact of straylight, recovering $\sim$0.14 mag. This would increase the aforementioned predicted upper limit to $\sim$900,000, a 13\% increase (all Galactic components increase by several million).  {\it Gaia}'s onboard software was updated in April 2015 to observe with adaptive AC widths. This new functionality has been successfully commissioned but depends very much on RVS calibrations and so is still being optimised.

 \vspace*{-0.6 cm}
\section{Summary}

{\it Gaia}-RVS is already the largest ever spectroscopic survey (5.4 billion spectra observed in its first year).  This means it will also be the largest ever survey of the Milky Way's halo.  It will provide $\sim$1 km~s$^{-1}$ precision radial velocities for $V < 12$ mag, which is the planned CU6 contribution to the second {\it Gaia} data release (2017, to be confirmed).  With {\it Gaia} astrometry, this will provide $\sim$10,000 {\it local field} halo stars (all Galactic components $\sim$2 million) in 6D.  RVS spectra will also be used by DPAC CU8 to derive abundances (Fe, Ca, Ti, Si), $T_{eff}$, log $g$, [M/H] and $A_{0}$ for $V < 12$ mag, ready for later data releases.  CU6-measured $v$sin$i$ for these stars means that {\it Gaia} will measure these $\sim$10,000 local field halo stars and the $\sim$2 million all-Galactic-component stars in a total of 15 dimensions.  

We present preliminary evidence that {\it Gaia}-RVS can provide $\sim$15 km~s$^{-1}$ precision end-of-mission radial velocities at $15 < V < 16$ mag.  At this precision and with {\it Gaia} astrometry, {\it Gaia} will provide $\sim$800,000-900,000 halo stars (all Galactic components $\sim$75-100 million) in 6D in the final {\it Gaia} catalogue (2022, to be decided).  When analysed with CU8-derived [M/H] from {\it Gaia}'s BP/RP spectra of these stars, {\it Gaia}'s chemo-6D-kinematic mapping out to $\sim$30-50 kpc from the Sun will revolutionise our understanding of the Milky Way halo's structure, origin and evolution.

\end{document}